\documentclass{elsart}
\usepackage{epsf}

\newcommand{\AmS}{{\protect\the\textfont2
  A\kern-.1667em\lower.5ex\hbox{M}\kern-.125emS}}
\begin{document}
\hspace*{\fill}\large HEPSY 99-2\linebreak
\hspace*{\fill}\large Feb. 1999~\linebreak
\begin{frontmatter}

\title{The CLEO-III Ring Imaging Cherenkov Detector} 

\author{R.J.~Mountain, M.~Artuso, R.~Ayad, F.~Azfar, A.~Efimov, S.~Kopp, }
\author{G.~Majumder, S.~Schuh, T.~Skwarnicki, S.~Stone, G.~Viehhauser}
\address{Syracuse University, Syracuse NY 13244--1130, USA}
\author{S.~Anderson, A.~Smith, Y.~Kubota}
\address{University of Minnesota, Minneapolis MN 55455, USA}
\author{E.~Lipeles}
\address{California Institute of Technology, Pasadena CA 91125, USA}
\author{T.~Coan, J.~Staeck, V.~Fadeyev, I.~Volobouev, J.~Ye}
\address{Southern Methodist University, Dallas TX 75275, USA}


\begin{abstract} 
The CLEO-III Detector upgrade for charged particle identification is discussed. 
The RICH design uses solid LiF crystal radiators coupled with 
multi-wire chamber photon detectors, 
using TEA as the photosensor, 
and low-noise Viking readout electronics.  
Results from 
our beam test at Fermilab 
are presented. 
\end{abstract}

\end{frontmatter}

\vspace{2cm}
\normalsize
\dotfill

{\em Invited talk by R. Mountain at ``The 3rd International Workshop on Ring Imaging Cherenkov
Detectors," a research workshop of the Israel Science Foundation, Ein-Gedi,
Dead-Sea, Israel, Nov. 15-20, 1998}
\newpage

\section{INTRODUCTION} 		
 %
The CLEO detector is undergoing a major upgrade (CLEO-III) in conjunction with 
a luminosity upgrade of the CESR electron-positron collider 
(CESR Phase-III) \cite{Kopp96,Artu98}.   
This upgrade will increase the luminosity of the machine 
by more than a factor of 10, 
to ${\pounds}=2\times 10^{33}$ cm$^{-2}$sec$^{-1}$, or
$\sim$20~fb$^{-1}$/yr, 
allowing unprecedented sensitivity to 
study CP violation in charged $B$ decays as well as 
the phenomenology of rare $B$ decay modes (with ${\rm BR}\approx 10^{-6}$). 
  %
Charged hadron identification is crucial in distinguishing 	
these decay modes. 
Typically, one wants highly efficient $\pi/K$ separation 
with 
mis-identification probabilities $\le10^{-2}$ 
over the full momentum range of secondaries 
from $B$-hadrons produced at the $\Upsilon(4S)$ resonance. 
Achieving this capability in modern particle detection 
has heretofore been elusive. 
 %
 %

We at CLEO believe that the best way to accomplish this task is to construct a 
Ring Imaging Cherenkov (RICH) Detector.

\section{GENERAL PRINCIPLES OF A PROXIMITY-FOCUSED RICH} 
 %
The CLEO-III RICH detector 
consists of three components: radiator, expansion volume, and 
photon detector.  
No focusing is used; this is called ``proximity-focusing'' 
\cite{t+j}.  
When an incident charged particle with sufficient momentum ($\beta>1/n$) passes 
through a radiator medium, 
it emits photons 
at an angle $\Theta$ via the Cherenkov effect; 
some photons are internally reflected 
due to the large refractive index $n$ of the radiator, and some escape.  
These latter photons propagate in a transparent expansion volume, 
sufficiently large to allow the Cherenkov cone 
to expand in size 
(as much as other spatial constraints allow). 
The photons are imaged by a two-dimensional pad detector, 
a photosensitive multi-wire chamber 
which records their 
spatial position. 
 %
The resulting images are portions of conic sections, 
distorted by refraction and truncated by internal reflection 
at the boundaries of media with different optical densities. 
Thus, knowing the track parameters of the charged particle and the 
refractive index of the radiator, one can reconstruct the 
Cherenkov angle $\Theta=\cos^{-1}(1/n\beta)$ 
and extract the particle mass.

This elegant and compact approach was 
pioneered by the Fast-RICH Group \cite{Arno92,Guyo94,Segu94}. 	


In order to achieve efficient particle identification with low fake rates, 
we set as a design goal a system capable of 
$\pi/K$ separation with 4$\sigma$ significance 
($N_\sigma=\Delta\Theta/\sigma_\Theta$) 
at 2.65 GeV/$c$, the mean maximum momentum for two-body $B$-decays 
at a symmetric $e^+e^-$ collider.  
At this momentum, the $\pi$-$K$ Cherenkov angle difference 
$\Delta\Theta=14.4$ mrad, which along with $1.8\sigma$ $dE/dx$ identification 	
from the central Drift Chamber, 
requires a Cherenkov angle resolution $\sigma_\Theta=4.0$ mrad per track. 
Using the relation 
$\sigma_\Theta=\sigma_{\Theta \rm pe}/\sqrt{N_{\rm pe}}$, 
we can establish   
round-number benchmarks for our design: 
a resolution of 14 mrad per photoelectron \cite{Efim95},   
and a photoelectron yield of 12 pe per track.

\section{BASIC DETECTOR DESIGN} 
 %
The overall RICH design is 
cylindrical, 
with compact photon detector modules at the outer radius and 
radiator crystals at the inner radius, forming thirty 12$^\circ$ sectors 
in azimuth. 
A schematic of the RICH is given in Ref.\ \cite{Kopp96}.  
The RICH resides between the central Drift Chamber and the CsI Calorimeter. 
The ensuing 
space budget constrains the detector to fit in 80--100 cm in radius, 
and 2.5 m in length (82\%\ of the solid angle). 	
The mass budget restricts the thickness to 12\%\ $X_{\rm o}$ 
to avoid significantly degrading the performance of the Calorimeter. 

However, the driving constraint of the design in many respects is 
the choice of Triethylamine (TEA) as a photosensor, 
which is both chemically aggressive and 
manifests a quantum efficiency in the VUV regime 
(135--165 nm). 
This greatly restricts the available materials: 
optical materials must be transparent in the VUV, and 
construction materials need to be 
chemically resistant to TEA 
and 
low outgassing. 

Each detector component and related design issues are discussed in turn. 


\section{CRYSTAL RADIATORS}
 %
The baseline design for the radiator is an 
array of individual planar LiF crystals\footnote{The 
 LiF and CaF$_2$ crystals are grown and polished 
 by OPTOVAC, North Brookfield, MA.}, 
each \mbox{$\sim 170 \times 170$ mm$^2$} 
and 10 mm thick, mounted on an inner carbon fiber cylinder. 
Due to the high refractive index of LiF ($n=1.50$ at 150 nm), 
a large number of 
Cherenkov photons are internally reflected, 
resulting in only a partial ring being imaged 
(about 1/3 of the initial cone).  
This is especially severe 
for incident track angles under 15$^\circ$,    
where the radiator would have to be tilted on the inner cylinder  
for the incident track to exceed this angle. 

\begin{figure}[t] 
  \begin{center}
       \centerline{\epsfxsize 2.50in \epsffile{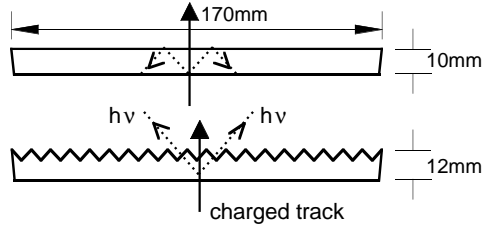}}
  \end{center}
\caption{
Azimuthal view of 
planar and sawtooth radiators with a normal incidence track. 
}
\label{fig:rad}
\end{figure}

In order to improve this situation, 
a novel radiator geometry has been implemented \cite{Efim95}, 
cf.\ Figure~\ref{fig:rad}. 
This ``sawtooth'' radiator, with its inner surface cut 
in profile to resemble the teeth of a saw, 
allows photons to cross the surface at near normal incidence. 
This reduces the photon loss by internal reflection and, 
as a consequence of the refraction angle, 
also the dominant chromatic error 
contribution to the resolution.  
Detailed 
Monte Carlo simulations \cite{Efim95} indicate that all performance parameters 
are better with 
a sawtooth radiator, and especially so at small values of incident track angle.  
Hence the central region of the detector will use 120 sawtooth radiators 
(47\% of the solid angle coverage), 
and the outer regions will use 300 planar radiators. 


\section{EXPANSION VOLUME} 
 %
The expansion volume is essentially empty space, 157 mm in radial distance, 
filled with pure N$_2$ gas. 
There are no support structures in the interior of the 
detector, which would be obstructions to photon propagation. 
However structural rigidity must be maintained with low mass, 
and so our mechanical design calls for 
aluminum end-flanges glued to the inner cylinder, 
and a detector module support frame, 
with 
reinforcing rib and box structures 
added to the modules. 


Importantly, 
UV photons will be lost 
if 
the expansion volume is not well-sealed from O$_2$ and H$_2$O contamination, 
and from any possible leakage from 
the photon detector volume. 
 %
Redundant gas seals are used in our design, 
and the N$_2$ will be exchanged at a 
flow rate sufficient to maintain high transparency. 

\section{PHOTON DETECTORS}   	
 %
\begin{figure}[t] 
  \begin{center}
       \centerline{\epsfxsize 3.00in \epsffile{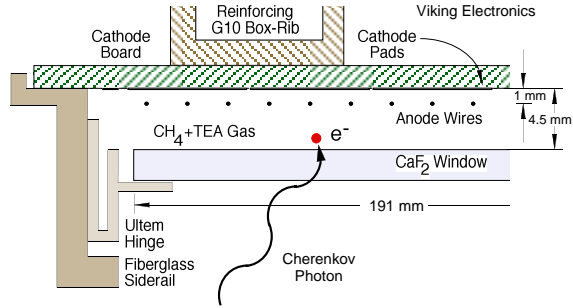}}
  \end{center}
\caption{
Schematic cross-section of a single photon detector module. 
A Cherenkov photon is shown propagating from the radiator. 
}
\label{fig:phdet}
\end{figure}
 %
The photon detector is a compact, photosensitive 
asymmetric multi-wire chamber, 
shown in Figure~\ref{fig:phdet}, 
filled with 
CH$_4$ carrier gas bubbled through liquid TEA at 15$^\circ$C 
(5.5\%\ vapor concentration). 
TEA has a peak QE of 33\%\ at 150 nm and a spectral bandwidth of 135--165 nm 
\cite{Segu94}.  
The detection sequence is: 
a photon passes through a thin UV-transparent 
CaF$_2$ window and is 
converted to a single electron 
by ionizing a TEA molecule. 
The 
single photoelectron drifts towards, 
then avalanches near, the 20 $\mu$m \O\ Au-W anode-wires, and 
induces a charge signal on the array of $8.0\times7.5$ mm$^2$ cathode-pads, 
providing a spatial point for the photoelectron.  
The design maximizes the photon conversion efficiency 
by having the wire-window gap be many photoabsorption lengths 
($\epsilon=99.9$\%, with $\ell_{\rm abs}=0.5$ mm at 150 nm), and 
maximizes the anode-cathode charge coupling ($C\approx 78\%$) 
by having the wire-pad gap be as small as practical (1 mm). 
Both may be optimized at once for a fixed thickness by making the chamber 
asymmetric.  

The CaF$_2$ window crystals 
must be deposited with metallized traces in order to act as field electrodes. 
 %
Moreover, 
the crystals 
are large ($191\times 308$ mm$^2$) and quite thin (2 mm) 
in order to minimize photon absorption and radiation length, 
and 
must be mounted with no mechanical stress to avoid inducing fractures. 
Our design utilizes Ultem hinges, to which the CaF$_2$ windows are glued, 
to relieve stresses.   
All construction materials in the chamber volume must be low outgassing and 
TEA-compatible.  
Extensive tests of TEA effects on 
various materials and adhesives of interest have been made.  
  %

\section{READOUT ELECTRONICS} 
 %
\begin{figure}[bt] 
  \begin{center}
       \centerline{\epsfxsize 3.00in \epsffile{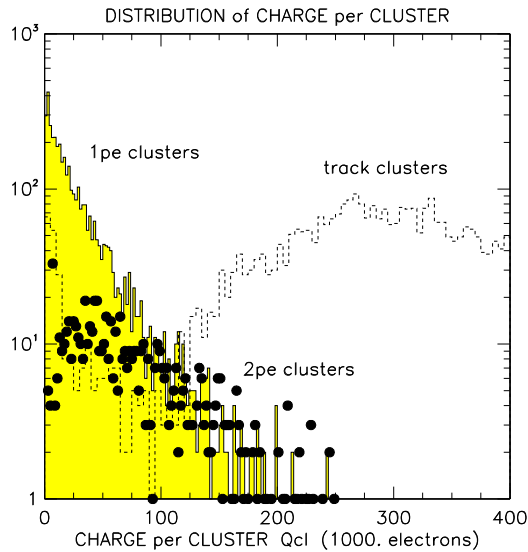}}
  \end{center}
\caption{
Distribution of charge per reconstructed cluster: 
for single-photoelectron clusters, 
two-photoelectron clusters, 
and charged-track clusters; 
on plateau, 
in CH$_4$+TEA (15$^\circ$C). 
Data from cosmic-ray tests. 
}
\label{fig:qpedist}
\end{figure}
 %
The choice of readout electronics is governed by the Furry 
(exponential) charge distribution of a single photoelectron avalanche, 
cf.\ Figure~\ref{fig:qpedist}, 
and also by the time allowed for the readout. 
The most likely charge is zero, but the tail is long, and so 
it is necessary to have analog front-end electronics 
with low noise and large dynamic range     
in order to maximize the photoelectron detection efficiency.  
The charge information is necessary in accurately determining 
the centroid of the photoelectron, 
as well as 
in disentangling the overlap of two nearby charge distributions.  
The latter requires high segmentation. 
Given the modest pad size over the large detector area, 
there are 230,400 total electronics channels. 
Occupancy is low ($<$1\%) and so sparsification is required.

The front-end signal processor is the 
Viking \cite{Nyga91} 
VA\_RICH chip, a new custom-designed 64-channel VLSI chip\footnote{This 
 Viking chip was designed and manufactured by IDE AS, Oslo, Norway.} 
incorporating these requirements, 
with 
measured rms noise 
${\rm ENC}=130 e^- + 9 C_{\rm det} e^-/{\rm pF}$ 
($\approx 150 e^-$ for modest $C_{\rm det}\approx 2$~pF), 
a shaping time of 1--3 $\mu$sec,    
and good 
linearity up to $\pm 4.5\times 10^5 e^-$ input.  
The chip has 
input protection, 
a preamplifier/shaper, 
sample \&\ hold circuitry, 
and a differential current output multiplexer. 
The analog signal travels over a 
6~m long cable from the detector to a 
VME databoard with receiver, 12-bit ADC and sparsifier.  

\section{BEAM-TEST SETUP} 
 %
\begin{figure}[t] 
  \begin{center}
       \centerline{\epsfxsize 3.00in \epsffile{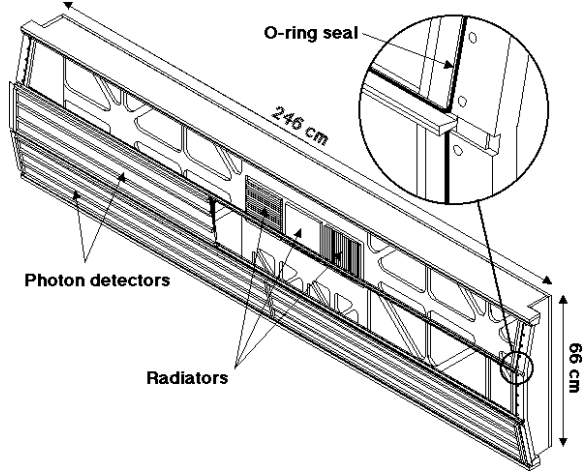}}
  \end{center}
\caption{
RICH beam-test box. 
Half of the middle chamber is cut away to show the radiators. 
The outer radiators were sawtooth radiators with the teeth 
arranged perpendicularly to each other.
}
\label{fig:btbox}
\end{figure}
 %
In order to test our understanding of the design and behavior of this detector, 
a comprehensive beam test was performed. 
 %
The first two completed photon detectors 
of the CLEO-III RICH 
were mounted on an aluminum box simulating the expansion volume, 
and equipped with one planar and two sawtooth LiF radiators 
(cf.\ Figure~\ref{fig:btbox}). 

The beam test \cite{Vieh98} 
was performed in a muon halo beam in the 
Meson East area of Fermilab, downstream of Experiment E866. 
The setup consisted of 
the RICH itself, 
its supporting gas and HV systems, 
trigger scintillators, 
and 
a charged-particle tracking system 
(2 MWPCs with 0.7 mm spatial resolution per station 
and combined $\sim$1 mrad track angle resolution). 
 %
Data was taken with the RICH box rotated at various 
polar and azimuthal angles 
to simulate the different incident track angles expected in CLEO-III.

Photon detector operation was stable during the three week beam test, 
running at a nominal gain of $4\times10^4$.  
The readout for the beam test consisted of 240 VA\_RICH chips and 8 VME 
databoards (for 15360 channels).  
After common-mode subtraction 
the remaining incoherent rms noise observed was $400 e^-$,  
providing an average signal-to-noise ratio for photoelectrons of 100:1.

\section{BEAM-TEST RESULTS}
 %
Photoelectrons 
(pe) are reconstructed 
by determining topological clusters of pads 
with pulse height 
above a 5$\sigma$ pedestal cut ($\sigma = 400 e^-$).  
 %
The overlap of multiple photoelectrons in a given cluster is 
disentangled by using the pulse height profile. 
The unbiased centroid is then found as the location of the photoelectron.  
(At operating voltage, 
pad multiplicities are 2.2 pads per cluster, 
and 1.1 pe per cluster.)
From each photoelectron position, 
the original photon trajectory 
is optically traced back 
through all media 
to the center of the radiator, and 
the Cherenkov angle is reconstructed. 
Charged track clusters are distinguished 
from photon clusters 
by total charge   
and by the number of pads in the cluster.

Data has been taken at a variety of track angles; 
in the following discussion 
only the datasets for the plane radiator at 30$^\circ$ track incidence and 
for the sawtooth radiator at 0$^\circ$ 
are considered in detail.  

\begin{figure}[t] 
  \begin{center}
       \centerline{\epsfxsize 6.00in \epsffile{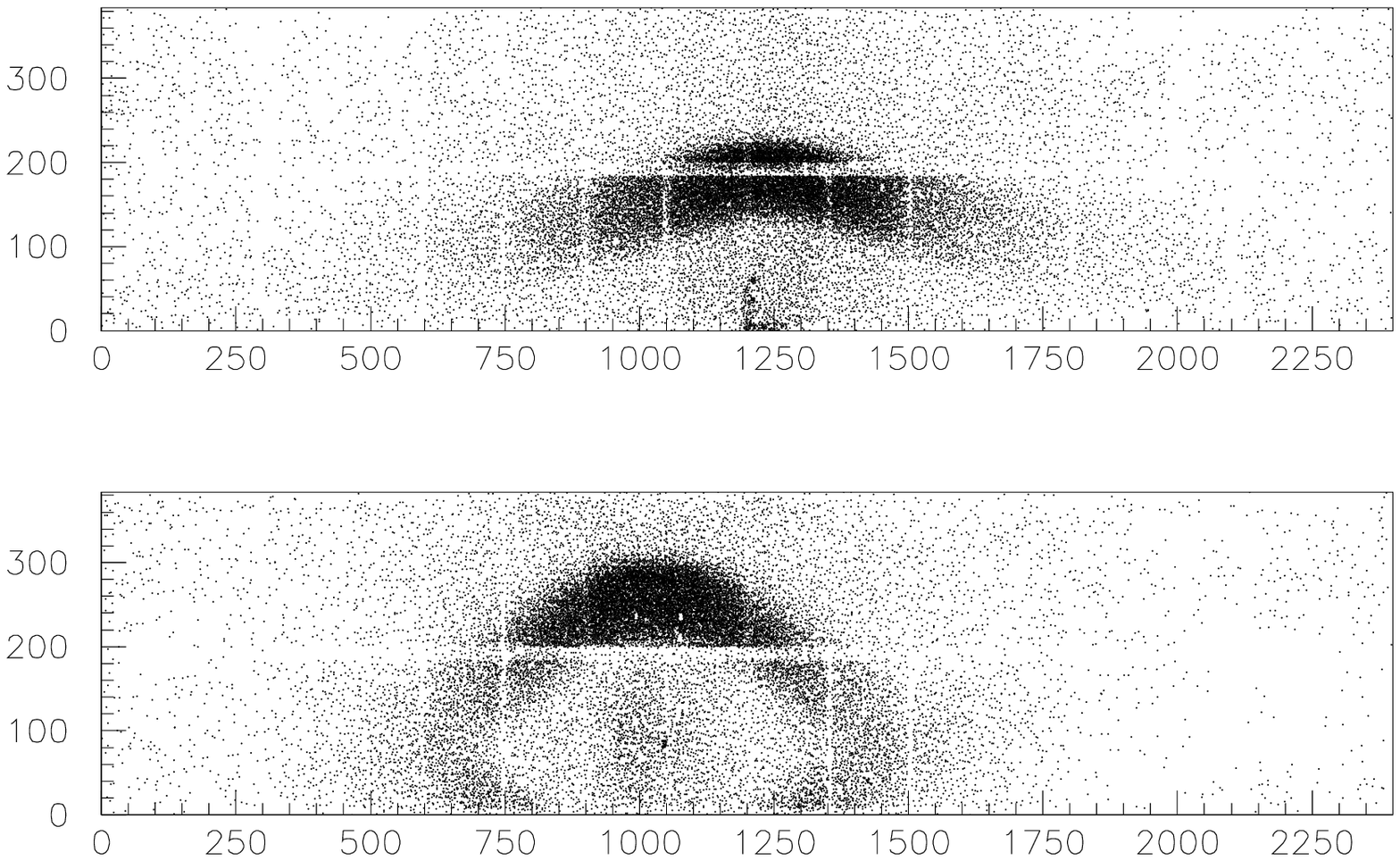}}
  \end{center}
\caption{
Cumulation of 
2489 Cherenkov images in (upper) 30$^\circ$ plane radiator dataset, and 
7434 images in (lower) 0$^\circ$ sawtooth radiator dataset.  
Units are mm. 
The bottom chamber contains the beam track, and 
is parallel to the radiator.  
Shadows of structural elements of the photon detectors can be seen.  
}
\label{fig:cum}
\end{figure}


Figure~\ref{fig:cum} 
shows a cumulative event display 
for all ring images in these two datasets.  
For the plane radiator one arc of the Cherenkov ring is 
visible, while for the sawtooth radiator two arcs in opposition 
are visible, 
with the lower one largely outside of the fiducial region of the detectors.  
Acceptance is lowered by this image truncation, 
and by mechanical transmission losses from construction elements 
in the detector. 
The acceptance for contained plane radiator images is 
the maximum realistic acceptance for a full RICH system, which is about 85\%;
the acceptance for sawtooth images is 
approximately 50\%\ in the two-sector beam test setup.



\begin{figure*}[t] 
\begin{center}
\begin{tabular}{ccc}
       \epsfxsize 2.00in \epsffile{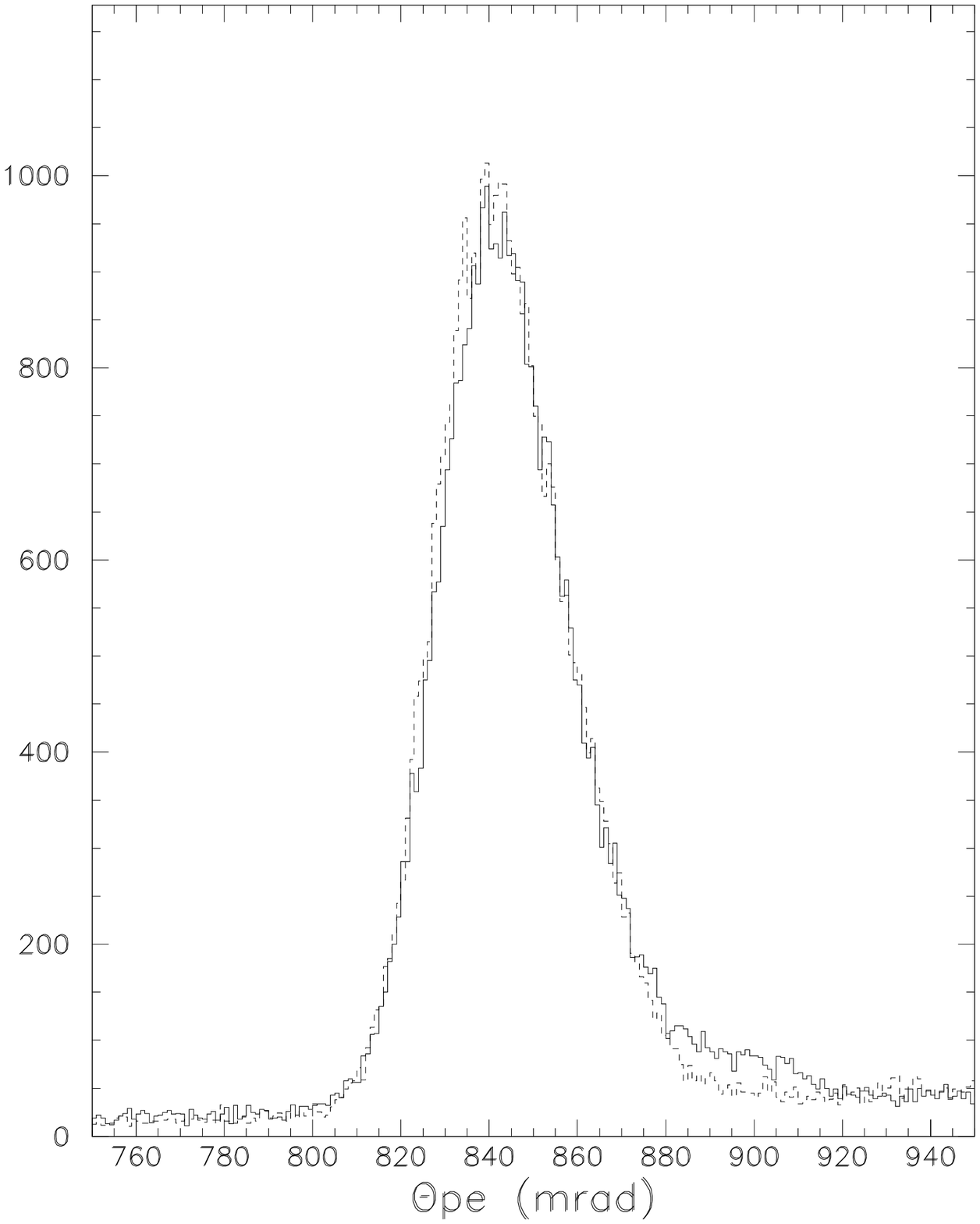} 
&      \epsfxsize 2.00in \epsffile{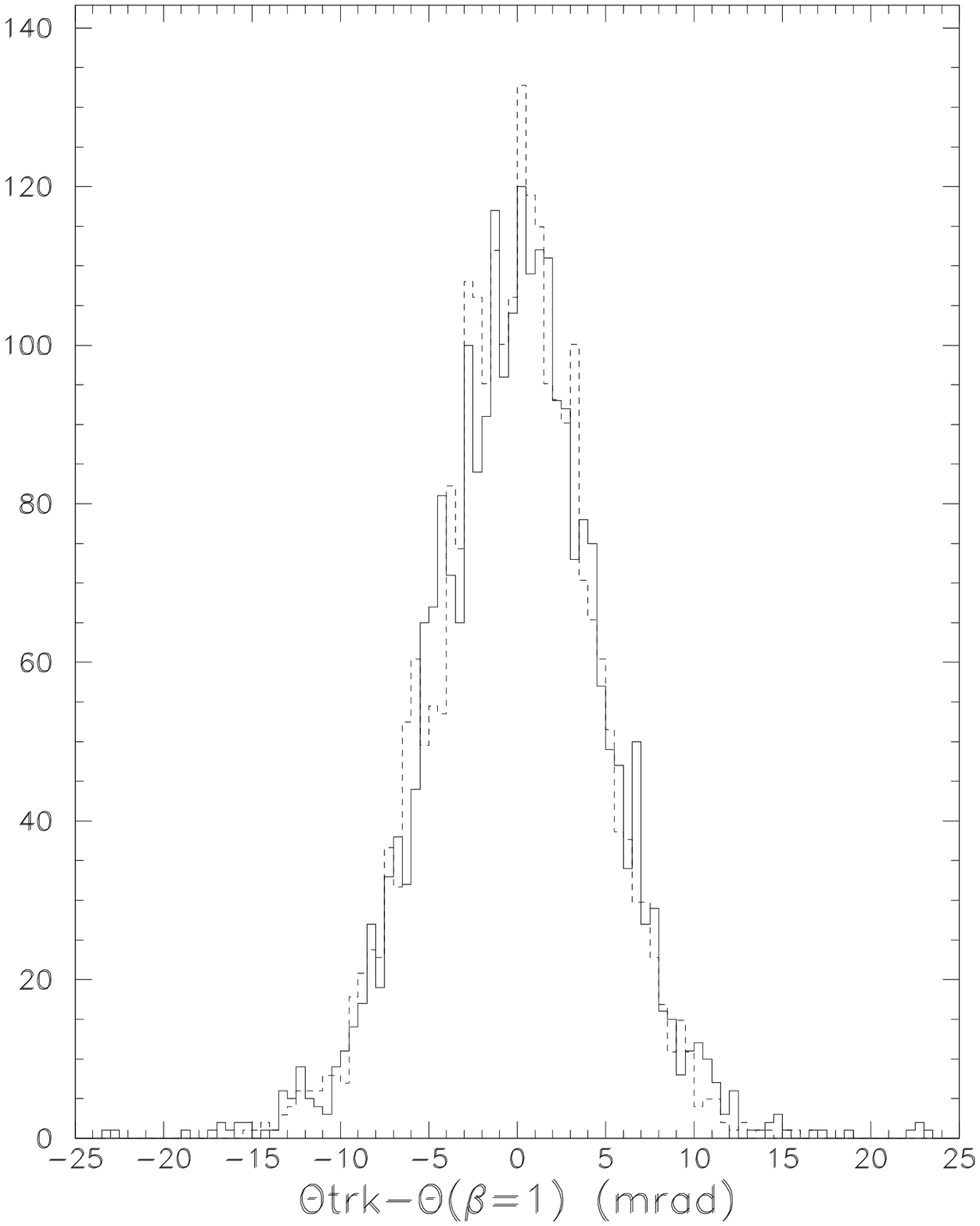} 
&      \epsfxsize 2.00in \epsffile{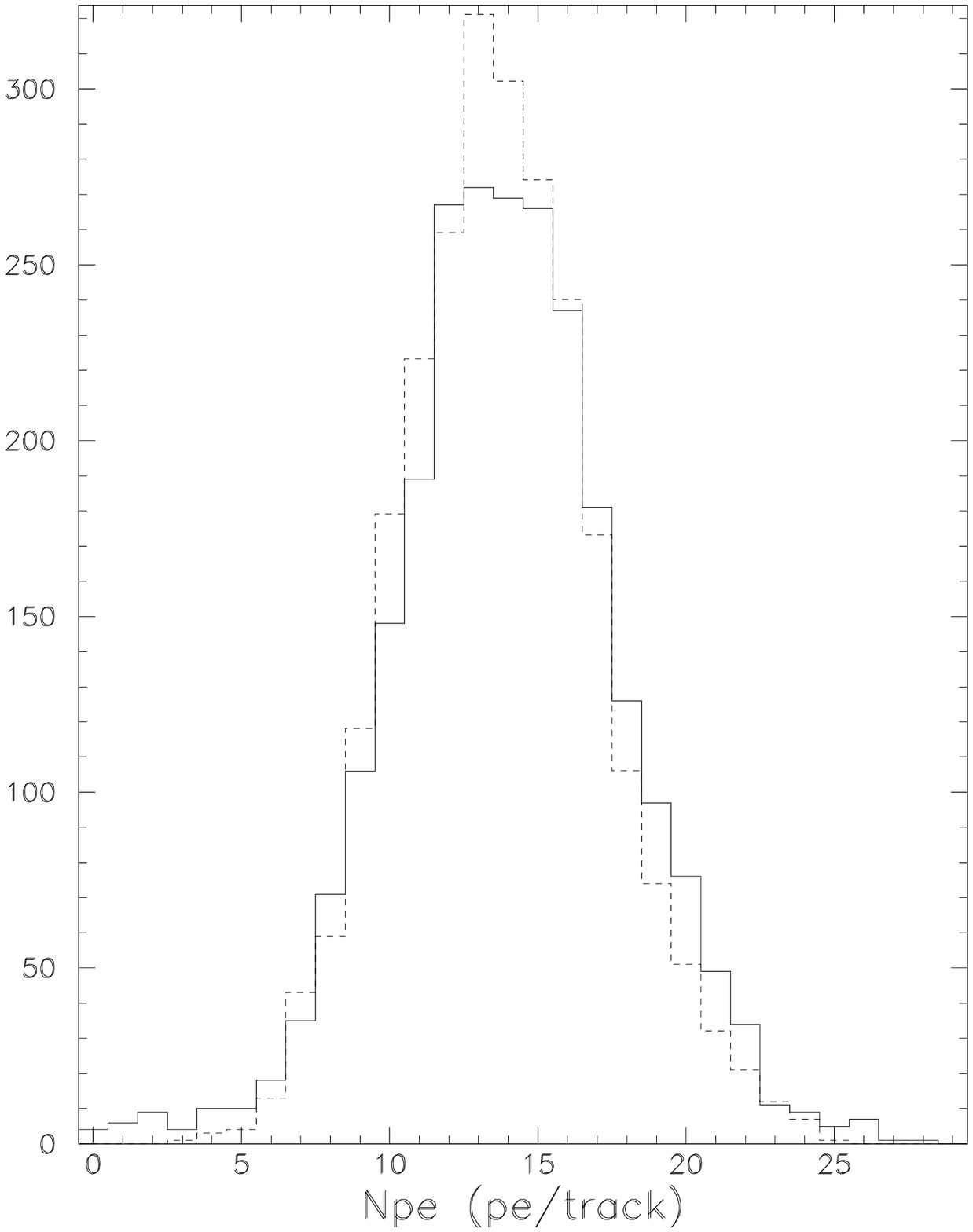}
\end{tabular}
\end{center}
\caption{
Plane radiator results. 
(a) Single photoelectron Cherenkov angle distribution; 
(b) Distribution of Cherenkov angle per track 
    shifted by angle of a high momentum muon ($\Theta_{\beta=1}$); 
(c) Photoelectron yield (pe/track). 
Solid line is for data, dashed line is for Monte Carlo. 
}
\label{fig:respl}
\end{figure*}

Results from the analysis of the 30$^\circ$ plane radiator dataset are shown 
in Figure~\ref{fig:respl}; 
only images confined to a single detector are used.  
The distribution of Cherenkov angle for single photoelectrons has 
an asymmetric tail and modest background; 
it is fit with a 
Crystal-Ball\footnote{The Crystal-Ball lineshape is 
 a Gaussian with an exponential tail at higher angles. 
 Resolution ($\sigma$) is extracted from the full-width at half-maximum.  
} lineshape plus polynomial background, 
yielding 
a single photoelectron Cherenkov angle resolution 
$\sigma_{\Theta{\rm pe}} = (13.2 \pm 0.05 \pm 0.56)$~mrad 
with a background fraction of 9.2\%\ under the image 
and a Monte Carlo estimate of 13.5~mrad. 
Errors quoted are first statistical then systematic, 
with the latter taken from two different fitting procedures, 
i.e., two methods of background estimation.  
This background is 
not electronic noise 
but rather it is 
principally due to out-of-time hadronic showers 
from an upstream beam dump; there will be no such background in 
the CLEO-III running conditions.  

The Cherenkov angle per track is found as the arithmetic mean 
of all photoelectrons in an image.  There is an 
image cut of $\pm3\sigma_{\Theta{\rm pe}}$ and 
a systematic alignment correction applied.  
The resulting distribution of Cherenkov angle per track 
is fit to a Gaussian, and 
gives the angle resolution per track 
$\sigma_{\Theta{\rm trk}} = ( 4.54 \pm 0.02 \pm 0.23 )$~mrad, 
which compares favorably with the Monte Carlo estimate of 4.45~mrad. 
The systematic error is estimated from 
the variation\footnote{This 
 variation has a number of root causes, 
 each at the few percent level: 
 the expansion volume transparency was monitored 
 to be above 95\%, 	
 there are variations in transparency over each radiator face, 
 etc.  
 Hence the systematic error is estimated to be at the 5\%\ level. 
} between different datasets taken at the same track angle, 
which are repeatable 	
to 5\%.  


The photoelectron yield 
$N_{\rm pe} = (12.9 \pm 0.07 \pm 0.36)$~pe per track 
is extracted from the area under the single photoelectron peak 
followed by background subtraction.  
Again 
systematic errors dominate and are given by different methods of 
background estimation.  
(Here the beam-test Monte Carlo makes no prediction for $N_{\rm pe}$ 
but rather uses the measurement as an input parameter.)  
This yield exceeds our benchmark of 12 pe/track.

\begin{figure*}[t] 
\begin{center}
\begin{tabular}{ccc}
       \epsfxsize 2.00in \epsffile{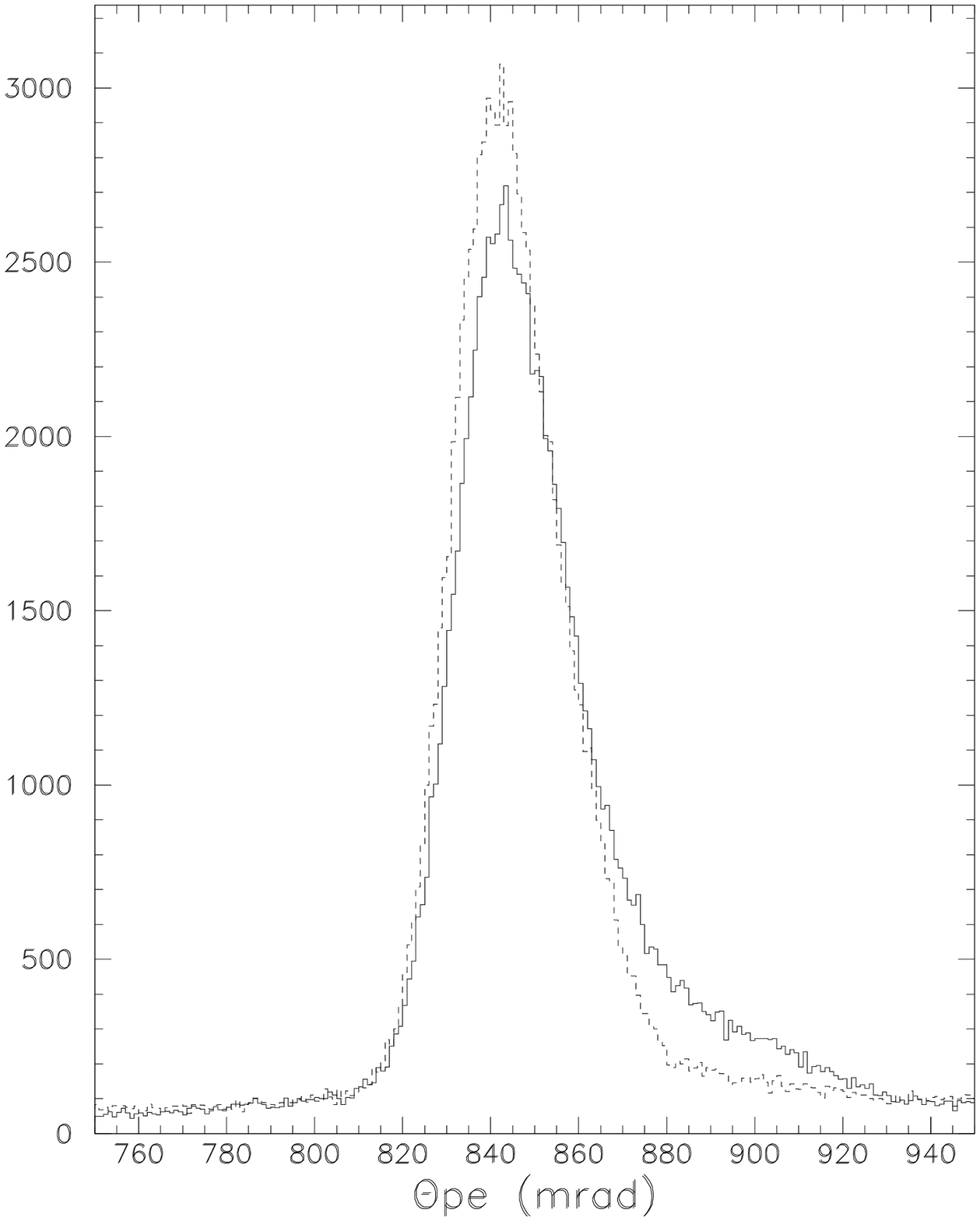} 
&      \epsfxsize 2.00in \epsffile{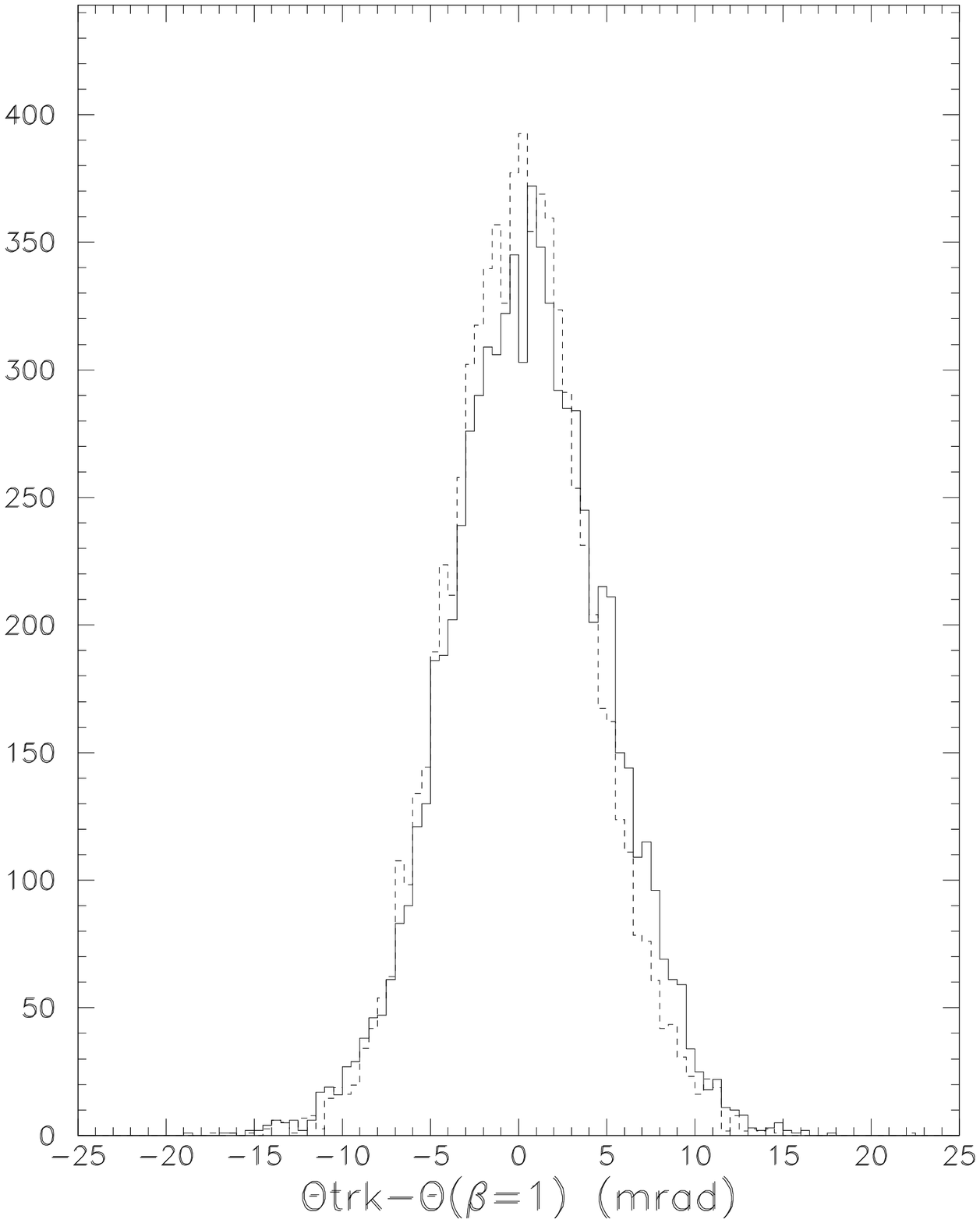} 
&      \epsfxsize 2.00in \epsffile{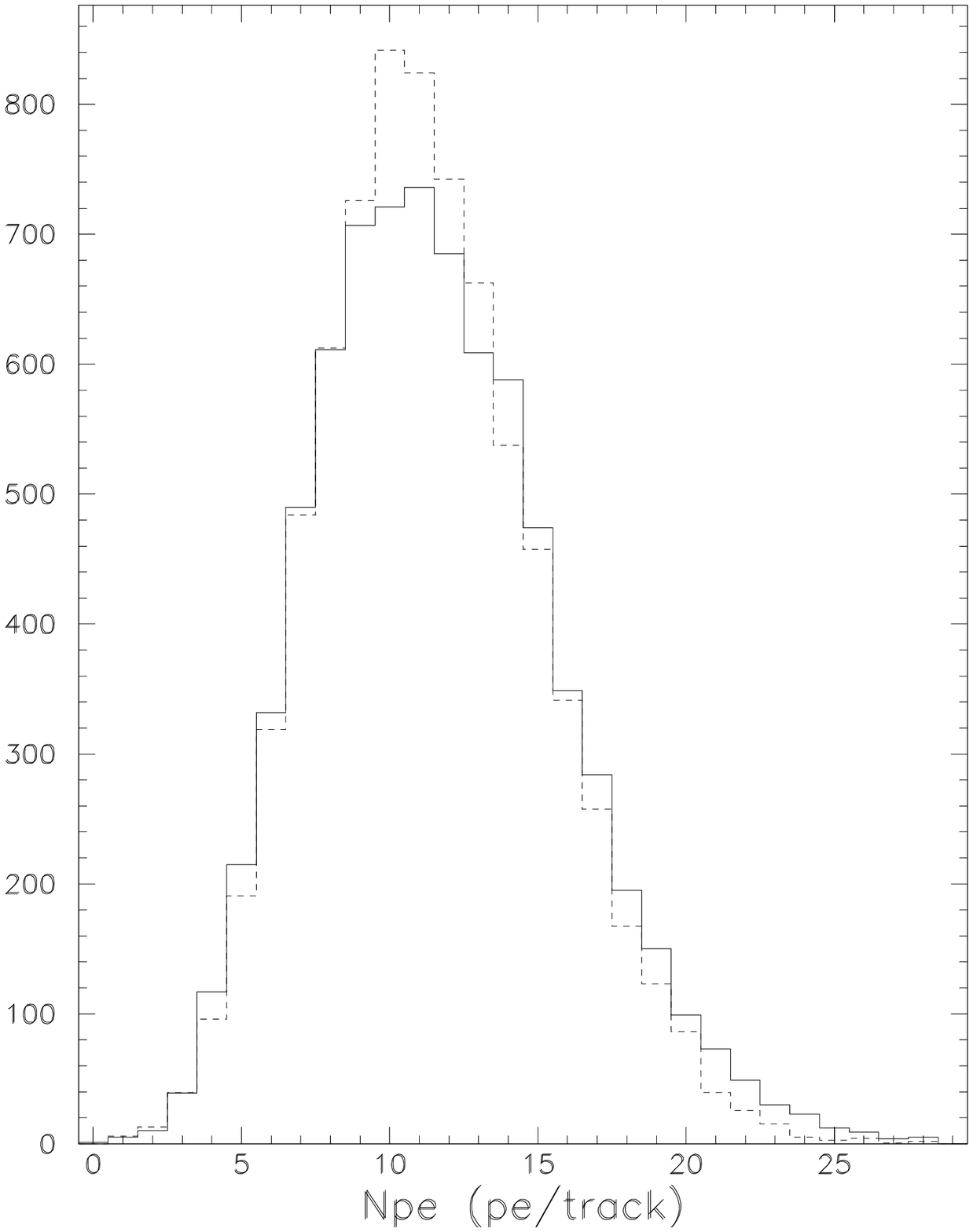}
\end{tabular}
\end{center}
\caption{
Sawtooth radiator results. 
(a) Single photoelectron Cherenkov angle distribution; 
(b) Distribution of Cherenkov angle per track 
    shifted by angle of a high momentum muon ($\Theta_{\beta=1}$); 
(c) Photoelectron yield (pe/track). 
Solid line is for data, dashed line is for Monte Carlo. 
}
\label{fig:resst}
\end{figure*}

Similar analysis for the 0$^\circ$ sawtooth radiator dataset, 
cf.\ Figure~\ref{fig:resst}, 
gives 
a single photoelectron Cherenkov angle resolution 
$\sigma_{\Theta{\rm pe}} = (11.7 \pm 0.03 \pm 0.42)$~mrad 
with a background fraction of 12.0\% 
(compared with 11.1~mrad from Monte Carlo), 
an angle resolution per track 
$\sigma_{\Theta{\rm trk}} = ( 4.49 \pm 0.01 \pm 0.22 )$~mrad 
(4.28~mrad from Monte Carlo), 
and 
a photoelectron yield 
$N_{\rm pe} = (10.4 \pm 0.04 \pm 1.0)$ pe/track, 
background subtracted.  
Adjusted for full 85\%\ geometric acceptance, 
$N_{\rm pe}$ becomes 18.8 pe/track.


\begin{figure}[t] 
  \begin{center}
       \centerline{\epsfxsize 3.00in \epsffile{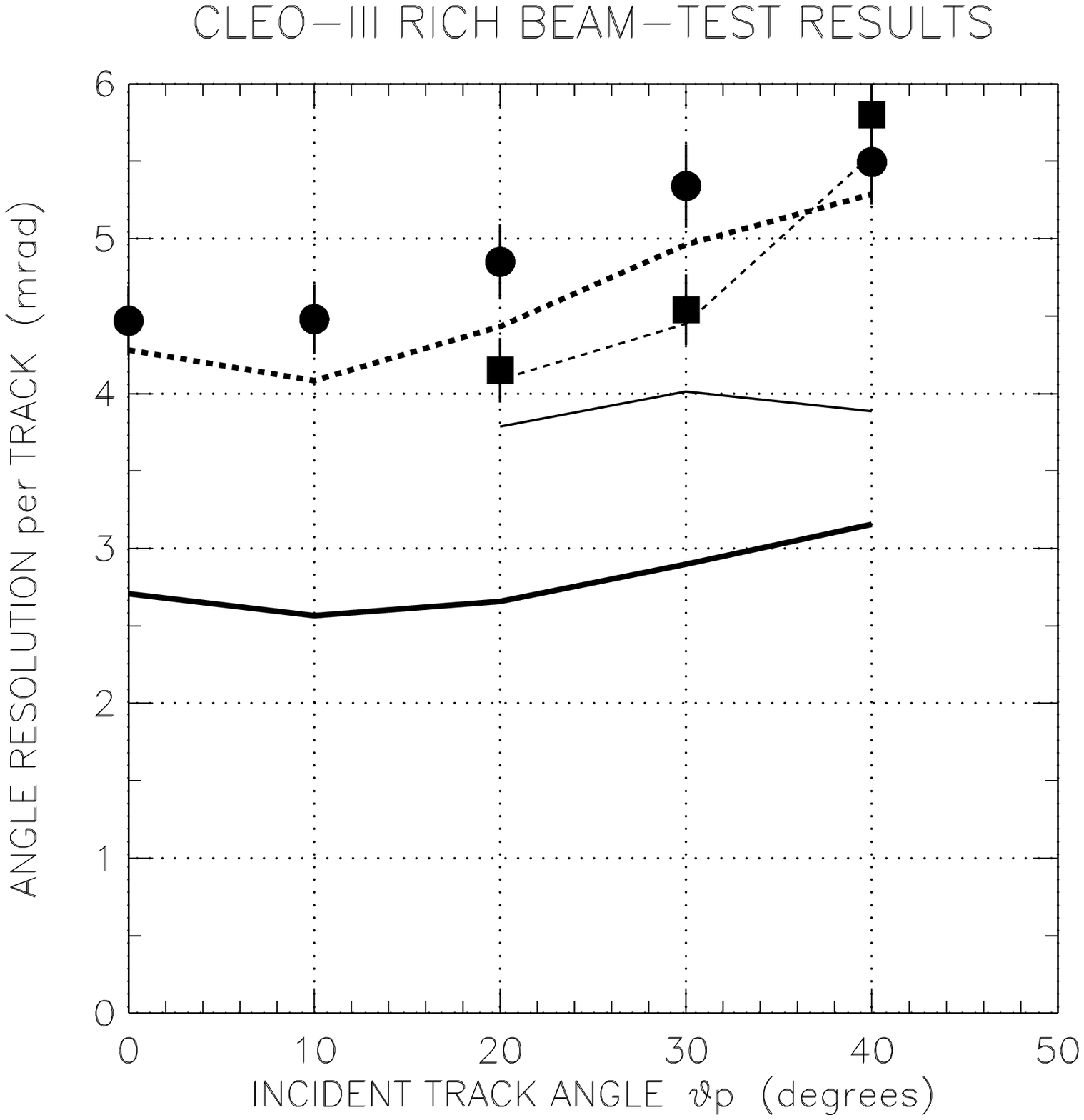}}
  \end{center}
\caption{
Summary of CLEO-III RICH beam-test results. 
Squares indicate plane radiator results, and 
circles sawtooth radiator results. 
The filled symbols represent results from beam-test data, 
dashed curves results from beam-test Monte Carlo, 
and 
solid curves results from the ``full acceptance'' extrapolation.  
}
\label{fig:ressum}
\end{figure}

Figure~\ref{fig:ressum} provides a summary of beam-test results 
from all datasets at all incident angles. 
The measured Cherenkov angle resolution per track from the plane radiator data 
(denoted by squares) 
increases as a function of the incident track angle 
due to the increase 
in emission-point error.\footnote{The 
  Cherenkov angle resolution per track is dominated 
  by chromatic and emission-point errors \cite{t+j}. 
  The chromatic error is larger at small track angles, 
  but they become comparable at large track angles. 
} 
The beam-test Monte Carlo simulation gives the light dashed curve 
in Figure~\ref{fig:ressum}, 
which represents the data well.  

However the per track resolution, e.g. 4.54~mrad at 30$^\circ$, is larger 
than that naively calculated by statistics, 
i.e. $13.2\, {\rm mrad}/\sqrt{12.9}=3.68\, {\rm mrad}$. 
 %
Monte Carlo studies indicate that 
the sources of the increased resolution 
are the MWPC tracking errors (the principal cause, 2.3~mrad at 30$^\circ$) 
and the beam background (1.2~mrad at 30$^\circ$).  
The tracking errors per se obviously cannot change with rotation of the RICH box, 
rather they 
effectively increase the emission-point error 
due to an incorrect track impact point on the radiator face, 
and hence become more prominent with track angle. 
In CLEO-III the background will be much reduced, 
and the tracking error contribution will be smaller 
yet still significant.

In order to estimate the 
ultimate performance of this RICH, 
an extrapolation was made based on the beam-test Monte Carlo. 
 %
The background and tracking errors 
are associated only with our beam test, so both 
were removed from the simulation for this study.  
The resulting photoelectron yield was then 
corrected for the geometric acceptance of the beam-test setup 
and scaled to ``full acceptance'', 
defined as 85\%\ of the solid angle covered 
by a cylindrical RICH.  
 %
The result of this ``full acceptance'' extrapolation for the 
per track resolution for the plane radiator 
is shown as the light solid curve in Figure~\ref{fig:ressum}, 
which is flat in track angle and 
below our benchmark of 4 mrad for CLEO-III.

The measured per track resolution from the sawtooth radiator data 
(denoted by circles in Figure~\ref{fig:ressum}) 
also increases with track angle, as expected \cite{Efim95}, 
again due to the increase in emission-point error. 
However the value of the measured per track resolution 
is larger than expected. 
This has several sources: 
acceptance, 
MWPC tracking errors, 
beam background, 
and 
sawtooth profile effects. 
Geometric acceptance is the largest contribution; 
it is approximately 50\%\ for all track angles. 
By naive statistical calculation this 
increases the per track resolution by 35\%. 
Monte Carlo studies show that 
tracking errors are the next largest contribution 
(e.g. 1.9~mrad at 0$^\circ$), 
and are exacerbated in this configuration because one of the arcs in the 
image is out of the detector fiducial.  
The beam background 
is approximately constant for all track angles 
(e.g. 1.3~mrad at 0$^\circ$). 
Sawtooth profile effects are defined as 
deviations of the real sawtooth radiator from an ideal sawtooth 
(e.g. rounding of the edges of teeth). 
However our simulation indicates that 
profile effects contributes little to the broadening of the resolution 
since mechanical imperfections are offset by a reduced transmission 
through the radiator in these same areas.

However even taking into account all these effects 
the beam-test Monte Carlo, 
which gives the heavy dashed curve in Figure~\ref{fig:ressum}, 
does not represent the data completely. 
It consistently underestimates the resolution, 
indicating that there are additional subtle systematic effects 
associated with the sawtooth radiator yet to be investigated. 


The ``full acceptance'' extrapolation 
for the per track resolution 
for the sawtooth radiator, 
the heavy solid curve in Figure~\ref{fig:ressum}, 
is flatter in track angle and 
closer to our expectations in value.

\begin{figure}[t] 
  \begin{center}
       \centerline{\epsfxsize 2.75in \epsffile{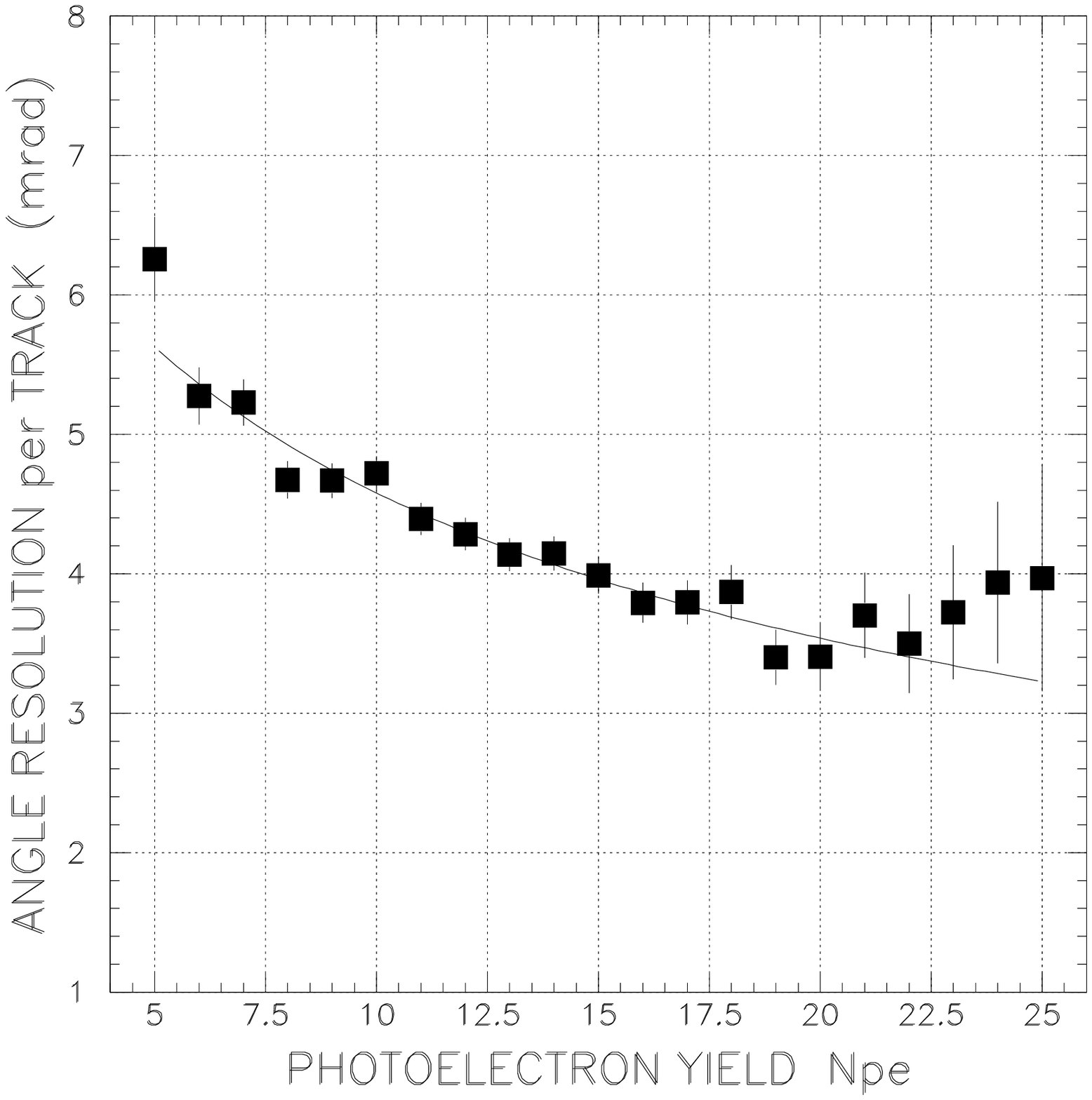}}
  \end{center}
\caption{The Cherenkov angle resolution per track 
as a function of photoelectron yield 
for sawtooth data at 0$^\circ$ track incidence.  
The curve is a fit to $A+B/\sqrt{N_{\rm pe}+C}$. 
}
\label{fig:mplot}
\end{figure}

A more sophisticated approach is made 
in Figure~\ref{fig:mplot}, which shows the 
Cherenkov resolution per track for the 0$^\circ$ sawtooth dataset 
as a function of photoelectron yield.  
One may read off the per track resolution at 
the measured yield of 10.4 pe/track 
and extrapolate to the expected yield of 18.8 pe/track, 
giving the result as 3.5~mrad.	
Since this curve is derived from the data 
it automatically takes into account 
statistical and systematic effects. 
Hence we have met our benchmark of 4 mrad.

To summarize, 
for both plane and sawtooth radiators 
unfolding the effects of background and tracking error gives 
angle resolution per track very close to the expected values. 
The expected CLEO-III RICH performance will fall somewhere 
between the beam-test Monte Carlo curve 
and the ``full acceptance'' curve in Figure~\ref{fig:ressum}. 
Clearly this is sufficient to meet our needs 
for CLEO-III.

\section{CONCLUSIONS and OUTLOOK} 		
 %
A beam test of the first two sectors of the CLEO-III RICH Detector
has been successfully carried out.  
The results obtained fulfill 
CLEO-III requirements for 4$\sigma$ $\pi/K$ separation, 
particularly a Cherenkov angle resolution of about 4 mrad. 

The CLEO-III RICH Detector is in the final phase of construction. 
At present 
85\%\ of the photon detectors have been built 
(with 40\%\ fully tested); 
all CaF$_2$ windows, 
78\%\ of the LiF planar radiators, and 
51\%\ of the LiF sawtooth radiators have been delivered; 
and
all readout chips have been acquired and tested. 
Completion  
and installation is expected in Summer 1999.

\section*{ACKNOWLEDGMENTS}
  %
We would like to thank Fermilab for providing us with 
the dedicated beam time for our test, 
the Computing Division for its excellent assistance, 
and our colleagues from E866 
for their hospitality in the beamline.



\begin{thebibliography}{9} 
 %
\bibitem{Kopp96} 
 S.E.~Kopp, Nucl.\ Instr.\ Meth.\ A384 (1996) 61. 
\bibitem{Artu98} 
 M.~Artuso, ``Progress Towards CLEO III'', 
 to be published in the Proceedings of the 
 XXIX International Conference on High Energy Physics, 
 hep-ex/9811031 (1998). 
\bibitem{t+j} 
 T.~Ypsilantis and J.~S\'{e}guinot, Nucl.\ Instr.\ Meth.\ A343 (1994) 30. 
\bibitem{Arno92} 
 R.~Arnold et al., Nucl.\ Instr.\ Meth.\ A314 (1992) 465. 
\bibitem{Guyo94} 
 J.-L.~Guyonnet et al., Nucl.\ Instr.\ Meth.\ A343 (1994) 178. 
\bibitem{Segu94} 
 J.~S\'{e}guinot et al., Nucl.\ Instr.\ Meth.\ A350 (1994) 430. 
\bibitem{Efim95} 
 A.~Efimov et al., Nucl.\ Instr.\ Meth.\ A365 (1995) 285. 
\bibitem{Nyga91} 
 E.~Nygard et al., Nucl.\ Instr.\ Meth.\ A301 (1991) 506. 
\bibitem{Vieh98} 
 G.~Viehhauser et al., Nucl.\ Instr.\ Meth.\ A419 (1998) 577. 
 %
\end{thebibliography}
\end{document}